\documentclass[aps,prl,twocolumn,groupedaddress]{revtex4}
\usepackage{epsfig}
\newcommand{\be}{\begin{equation}}
\newcommand{\ee}{\end{equation}}
\newcommand{\ba}{\begin{eqnarray}}
\newcommand{\ea}{\end{eqnarray}}
\newcommand{\pr}{{\rm pr}}

\begin{document}

\title{Some comments on computational mechanics, complexity measures, and all that}

\author{Peter Grassberger}

\affiliation{JSC, FZ J\"ulich, D-52425 J\"ulich, Germany}

\date{\today}

\begin{abstract}
We comment on some conceptual and and technical problems related to computational mechanics,
point out some errors in several papers, and straighten out some wrong priority claims. We 
present explicitly the correct algorithm for constructing a minimal unifilar hidden Markov model 
(``$\epsilon$-machine") from a list of forbidden words and (exact) word probabilities in a stationary
stochastic process, and we comment on inference when these probabilities are only approximately 
known. In particular we propose minimization of forecasting complexity as an alternative basis for 
statistical inference of time series, in contrast to the traditional maximum entropy principle.
We present a simple and precise way of estimating excess entropy (aka  ``effective measure 
complexity". Most importantly, however, we clarify some basic conceptual problems. In particular, 
we show that there exist simple models (called ``totally recurrent graphs") where none of the nodes 
of the ``$\epsilon$-machine" (the ``causal states") corresponds to an element of a state (or history) 
space partition. 
\end{abstract}
\maketitle

\tableofcontents

\section{Introduction}

During the last 25 years or so, there has been published a series of papers on minimal hidden 
unifilar Markov chains (mUHMC) \cite{footnote1}.
running under the heading of ``computational mechanics" and 
stressing aspects like optimal forecasting and estimates of structural complexity
\cite{Crutch-young1,Crutch-young2,Crutch1992,Crutch-feld97,Hanson-crutch,Crutch-shalizi,Shalizi-crutch,
Perry-binder,Crutch-feld03,Feld-crutch03,Shal-Shal-crutch,Shal-shal,Shal-klin-hasl,Crutch-ellison-mahoney,
Ellison-mahoney-crutch,Mahoney-ellison-crutch,Wiesner-rieper-vedral,Loehr,Haslinger-klinker-shalizi,
Mahoney11,Crutch12}. 
The first and seminal paper in this series was \cite{Crutch-young1}. Like many of its 
followers it is characterized by ingenious new nomenclature (``$\epsilon$-machine", 
``morph", ``causal state", ``statistical complexity", ...) for concepts that had been in use previously
under different names -- as well as by verbosity, and heavy mathematical jargon, and 
several crucial errors.

In this comment, I want to point out that: 

\begin{itemize}

\item Most of the concepts and constructs supposedly introduced in \cite{Crutch-young1}
had been introduced in \cite{Grass86}, including mUHMC's (``$\epsilon$-machines") , their nodes 
(``causal states") and forecasting complexity FC (``statistical complexity"; see also 
\cite{Grass86a,Grass87,Zambella,Grass88,Grass89,Grass89b}).
The principal author of \cite{Crutch-young1}, Jim Crutchfield, had been aware of these, since we 
had been together at a workshop in 1987 where I gave a talk and we had numerous 
discussions, in which I explained him in detail the properties of the FC and its relationship 
to other complexity measures like excess entropy {\bf E}. In particular, the fact that the 
inequality FC $\geq {\bf E}$ is usually not saturated is not due to Crutchfield {\it et al.},
in contrast to what is e.g. claimed in \cite{Mahoney-ellison-crutch}.  

\item The mathematical level of rigor in \cite{Crutch-young1} and in many of its 
follow-ups is not higher than in \cite{Grass86,Zambella,Grass88}, although the later 
papers were written in a much less formal style. In particular, it is {\it not} true that 
\cite{Crutch-young1} was needed to give ``operational definitions" to some of the 
concepts introduced in \cite{Grass86}, as claimed in \cite{Haslinger-klinker-shalizi}.
How can one give exact formulae and numerical values, as in \cite{Grass86,Grass86a,Zambella,Grass88}, 
without proper operational definitions? Yet, the same claim is also made in \cite{Shalizi-crutch}, 
page 871.

Quite to the contrary, \cite{Crutch-young1} made a number of conceptual and technical 
errors, some of which still are perpetuated in the computational mechanics literature 
\cite{footnote0}.

\item All agglomerative algorithms for finding (mUHMC)
(``$\epsilon$-machines") and their nodes (``causal states") given in 
\cite{Crutch-young2,Crutch1992,Perry-binder}
are incorrect. The (very simple) correct algorithm was briefly mentioned in \cite{Zambella}, 
p. 293, and will be described in detail below. As a consequence, it is not true that the divisive
algorithm of \cite{Shal-Shal-crutch,Shal-shal} is the only efficient and practical algorithm 
available.

\item If transient states are retained in the mUHMC, then the set of nodes does {\it not} in 
general correspond to a {\it partitioning} of the space of past histories, but to a {\it 
covering}, in which (at least) some histories are represented by several nodes \cite{Grass88}. 
This is particularly obvious for the start node, which corresponds to all possible histories.
But as shown in \cite{Zambella} and discussed in detail below, there are simple systems where 
{\it none} of the nodes correspond to an element of a partitioning.
It makes e.g. some of the proofs in \cite{Shalizi-crutch} technically wrong, although this could 
presumably fixed easily. In any case, as we shall see later, it makes the entire concept of 
$\epsilon$-machines ill defined in many non-trivial cases. 

\item In recent papers by Crutchfield {\it et al.}, transient parts of 
mUHMCs are never discussed -- maybe because they became aware of this problem. 
This is a pity, since they would simplify several problems.
In \cite{Crutch-ellison-mahoney,Ellison-mahoney-crutch,Mahoney-ellison-crutch} it is e.g. 
claimed that the only efficient algorithm for computing the excess entropy {\bf E} 
(called effective measure complexity (EMC) in \cite{Grass86}) is via $\epsilon$-machines for 
forward {\it and} backward processes. Instead, a very simple and efficient algorithm for 
computing {\bf E} from an mUHMC {\it including its transient} can be given by following remarks
on pp. 292-293 of \cite{Zambella}.

\item In parallel to this, there was increasing interest in measures of structural complexity.
Some of this activity has justly been criticized in \cite{Crutch-f-s}. Much more interesting 
are the attempts to define structural complexity in \cite{Gell-Mann-lloyd,Lloyd-gell-mann}
(see also \cite{Ay-mueller-szkola}).
What is missed in several recent papers is that the basic concept for defining structural 
complexity in \cite{Gell-Mann-lloyd} is 
the same as in \cite{Grass86,Grass89} and in \cite{koppel,atlan}. Indeed, the similarity
of names used by Gell-Mann \& Lloyd (``effective complexity") and in \cite{Grass86} (``effective 
measure complexity") suggests that these names were not chosen independently. In particular,
the idea that a structural complexity should be associated not to a single object but to an 
{\it ensemble} was not invented in \cite{Gell-Mann-lloyd} (as claimed in \cite{Ay-mueller-szkola}), 
nor by Dennett \cite{Dennett} (as claimed in \cite{Crutch-shalizi}), but was a main conclusion 
drawn explicitly in \cite{Grass86} and was at least implicit in \cite{koppel,atlan}.

\end{itemize}

In the following, I will discuss these points and some others in more detail.

\section{Measures for Structural Complexity}

After it had been pointed out (most vigorously by Atlan \cite{Atlan79}, and later by Huberman 
{\it et al.} \cite{Hogg}) that ``complex" systems should be neither ordered nor random, but in between,
it was clear that Kolmogorov ``complexity" was not the right indicator for what scientists would 
intuitively call complex behavior. I was unaware of R. Shaw's definition of ``stored information"
\cite{Shaw}, but was mostly motivated by Wolfram's treatment of the complexities of formal ``languages"
produced by cellular automata \cite{Wolfram}. One might have thought that the theory of hidden 
Markov chains would also have been a good starting point, but for whatever reasons it was not.
Three things became soon obvious: 

\begin{itemize}
\item a satisfactory definition of complexity needs to take into 
account probabilities, in contrast to the purely algorithmic approach dominating then, and best 
illustrated in the book by Hopcroft \& Ullman \cite{hopcroft}. 
\item In order to avoid the trap that 
a random object is `complex', because it is hard to describe in all its details, one has to 
attribute ``complexity" somehow not to individual objects, but to ensembles. After all, an 
ensemble of completely random objects is very easy to describe. 
\item There is a close connection between the complexity of strings and difficulties associated
with forecasting them. And there is a clear distinction between the degree to which a forecast 
is {\it possible} on the one hand, and the {\it difficulty} of making an optimal forecast.
A completely random sequence of zeroes and ones is impossible to forecast, but the {\it best}
forecast is trivial: It is a mere guess.
\end{itemize}

Within Kolmogorov-Chaitin algorithmic information theory, the last point was driven home
by Bennett \cite{Bennett} with the concept of {\it logical depth}: While algorithmic 
complexity measures the {\it amount} of information needed for constructing an object (i.e., for 
a symbol string, how much cannot be predicted), logical depth measures the {\it effort} needed 
to actually construct it from the most concise program (i.e., to actually perform the 
best possible prediction). 

Let us now discuss the {\it forecasting complexity} FC
(called ``true measure complexity" in \cite{Grass86}; the name forecasting complexity was 
proposed in \cite{Grass88,Zambella,Grass89b}, and is more descriptive than the name ``statistical 
complexity" introduced much later in \cite{Crutch-young1}). Its technical definition is easy 
for discrete stationary processes in discrete time. For any mUHMC which describes the 
process, one defines $C$ as the entropy of the stationary probability distribution on the 
(hidden) nodes of the Markov graph. The FC is then defined by using the {\it minimal} graph.
Technically, the minimal graph was obtained by adapting a well known algorithm for {\it topological}
Markov chains \cite{hopcroft} which works by joining nodes with the same forecasting ability. 
This gives the graph with the smallest number of nodes, but it was checked in all cases that
this graph has also the smallest value of $C$. 

Notice that since FC is the difficulty (per letter) of the optimal forecast, and the latter allows
to construct the entire sequence with minimal information (e.g., by means of arithmetic coding
\cite{Arith-code}), FC is indeed a realization of logical depth stationary processes.

Notice also that neither \cite{Grass86} nor \cite{Crutch-young1} contain a proof that the minimal
graph leads indeed to the minimal $C$, and to my knowledge such a proof is still lacking 
\cite{footnote9}. But 
while this problem was pointed out in \cite{Grass86}, it was simply missed in \cite{Crutch-young1}.

In \cite{Grass86}, no attempt was made to infer a mUHMC from data, for reasons detailed below. 
In lack of such a model the FC is not directly observable from empirical data. 
In order to alleviate this problem at least partially,
a lower bound of it was introduced which can be estimated directly from observations. This 
is {\bf E} (called EMC in \cite{Grass86}). It was defined in terms of block
entropies, and it was shown to have also an interpretation as mutual information between 
future and past (albeit in somewhat obscure words: ``the minimal information that would have
to be stored for optimal predictions if it could be used with 100\% efficiency"; in 
\cite{Feld-crutch03}, this discovery is attributed to much later papers 
by W. Li \cite{Li} and themselves \cite{Crutch-feld03}. Furthermore, it was shown that the 
inequality $ {\bf E} \leq FC$ is in general 
a strict one, i.e. equality between the two is rather the exception than the rule. In 
\cite{Ellison-mahoney-crutch} this inequality is attributed to \cite{Shalizi-crutch}. But in 
\cite{Crutch-young1} (page 107), instead of an inequality between ${\bf E}$ and $FC$ a ``simple
proportionality" was claimed.

Crutchfield {\it et al.} (see e.g. \cite{Crutch-young1,Crutch1994,Crutch-feld03}) claim that 
excess entropy was first introduced by 
Crutchfield \& Packard in \cite{Crutch-packard}. There is indeed a quantity with this name 
introduced in that paper (page 213), but it designates something completely different in 
a completely different context: it gives the amount by which the apparent entropy of a 1-d
map is increased by added external noise.

\section{Reconstruction of minimal UHMCs from exact word probabilities}

Assume we are given a list of all forbidden words in some formal language over a finite alphabet
$\cal{A}$. We will assume that any word is allowed which does not contain a subword in this list, 
and we will assume stationarity. If this list is finite, then the language is 
regular \cite{hopcroft}, and there exist well known algorithms for constructing the corresponding 
accepting deterministic automaton \cite{hopcroft}. The latter consists of a directed finite 
graph with a ``start" node and links labeled by letters $a_i\in \cal{A}$.
All links emanating from one node are labeled by different letters, so that any allowed word 
corresponds to a unique walk starting at the start node. Forbidden words would require absent
outgoing links, and are thus recognized by the fact that they do {\it not} correspond to such
walks. Furthermore we assume that there is at least one outgoing link from each node, so that
the language contains infinitely long words.

In the present case we are interested in a generalization where probabilities are associated to 
every allowed word, where the alphabet might be infinite (but still countable), and where there 
might be an infinite number of forbidden words. In this case the accepting graph might be infinite, 
and the probability distribution on the words induces a set of transition matrices between the 
nodes: $T^{(a)}_{ki}$ is the probability that a walker located at node $i$ will go to node $k$ 
and emit letter $a$. The resulting graph with attached probabilities is then a mUHMC. Under 
suitable conditions (e.g. when the process is ergodic and for finite
graphs, but also for many infinite graphs) there will be a unique stationary measure which is 
used, e.g., to define the entropy $C$ mentioned above.

Assume now that only a finite number of forbidden words and of word probabilities are known, but 
they are known {\it exactly}. For simplicity we assume also that, together with the probability
$\pr\{a_1 \ldots a_k\}$ we also know the probabilities for all subwords. We do {\it not} assume that
all word probabilities up to a fixed word length $k$ are known, i.e. as in written natural languages
we might know some branches of the suffix tree to a larger depth than others. In this case we cannot 
hope to construct the full mUHMC, but we can hope to construct the {\it simplest} UHMC compatible 
with the data.

The algorithm for doing so is indeed such a straightforward generalization of the algorithm for 
finding the smallest accepting deterministic automaton in the non-probabilistic case \cite{hopcroft},
that no details were given in \cite{Grass86}, and only a sketch was given in \cite{Zambella}.
It first constructs a non-minimal graph which represents all data exactly and allows all words that 
are not explicitly forbidden, and then minimizes it in a second step by identifying equivalent nodes.
On the other hand, a number of attempts were made in \cite{Crutch-young1,Crutch-young2,Crutch1992,Perry-binder}
which were all unsuccessful in allowing too many words, leading to non-minimal graphs, or not using 
the data optimally. This has lead in \cite{Shal-Shal-crutch,Shal-shal} to the claim that `agglomerative'
algorithms like the above are not competitive with their `divisive' one. We have not made yet any comparative 
test, but at first sight we cannot see any basis for this claim. In all test cases were we used 
either the divisive algorithm described below \cite{footnote2} or the agglomerative one described in 
\cite{Zambella}, both performed flawlessly. 

If the probabilities for all words of length $\leq k$ are given, the algorithm constructs a Markov model
of maximal order, i.e. of order $k-1$. In contrast, the algorithm described in
\cite{Crutch-young2} does not in general yield the Markov model of maximal order. Correspondingly,
many of the graphs given in \cite{Crutch-young1,Crutch-young2} are wrong (as, e.g. Fig.~1a in
\cite{Crutch-young1} and Figs.~8 and 9 in \cite{Crutch-young2}). The correct version of Figs.~8
in \cite{Crutch-young2} had been given before in \cite{Grass88}). Figure 1a in \cite{Crutch-young1} 
(for the logistic map at the Feigenbaum point) differs from the corresponding figure in \cite{Grass88} 
for two reasons. Apart of the problems discussed above, it disregards transients and accepts only 
sequences obtained from orbits on the attractor (while Fig.~4 in \cite{Grass88} accepts all trajectories). 
The correct graph for this problem would be the one given in Fig.~1.

\begin{figure}
\begin{center} 
\psfig{file=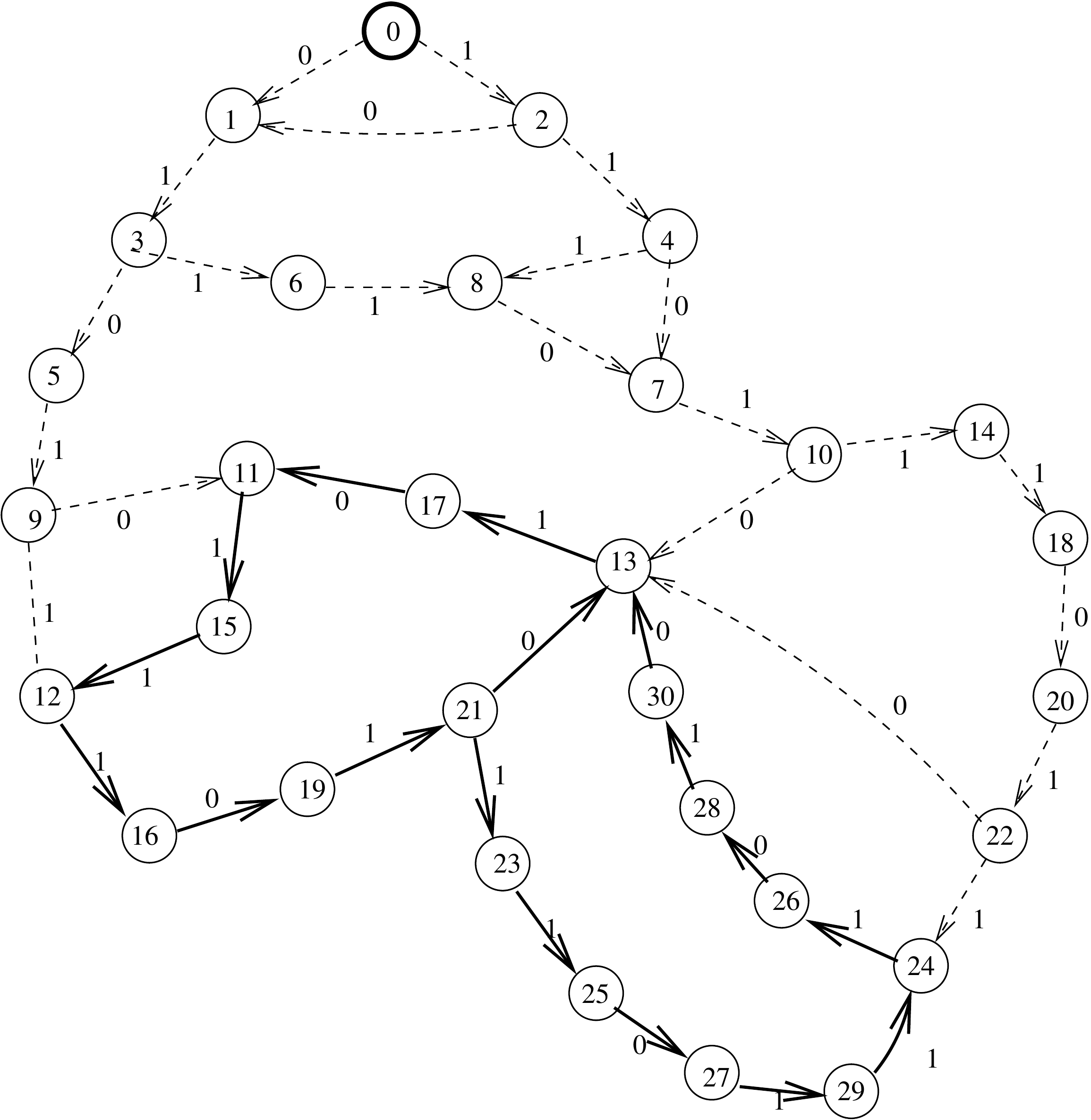,width=8.3cm, angle=0}
\caption{Minimal deterministic (unifilar) graph which accepts/produces all 0/1 sequences 
   on the Feigenbaum attractor up to length 16. Solid lines correspond to the recurrent part, dashed ones to 
   the transient. Notice that this is also the minimal graph which predicts
   their probabilities correctly, since each non-trivial branching ratio is 1:1 except the one at start,
   where $\pr\{0\}:\pr\{1\} = 1:2$. Compare this graph 
   (with 15 transient and 16 recurrent nodes) to the graph shown in \cite{Crutch-young1,Crutch-young2}
   which has 47 recurrent nodes.}
\end{center}
\end{figure}

The actual algorithm is best explained with an example. A typical set of data for a binary process might be 
\ba
   \pr\{00\}  &=& 0,\;\;\; \pr\{01\}  = \pr\{10\} = 1/4,\\
   \pr\{110\} &=& 1/6,\;\;\; \pr\{111\} = 1/3,
\ea 
from which the remaining probabilities for word lengths $\leq 2$ can be inferred by means of the Kolmogorov 
consistency relations. From this we construct a rooted binary ($k-$ary for alphabet size $k$) tree in which 
each listed word corresponds to a path from the root to a leave, and where we also note the conditional 
probabilities for each letter, conditioned on the previous letters in the word. The resulting tree is shown 
in Fig.~2a.

The next task is to transform this tree into a graph without any leaves, by identifying each leaf with a 
suitably chosen internal node. Remember that each leave corresponds to an allowed word. After the 
transform, its daughters 
will be those nodes that are reached by extending these words. Let us assume that the word leading from the 
root to leaf $w$ is $(a_1,a_2,\ldots a_k$). Obviously, we have no information about the extensions 
${\bf a} \equiv (a_1,a_2,\ldots a_k,a_{k+1})$ for either $a_{k+1}=0$ or $a_{k+1}=1$, but we might have 
information about $(a_2,\ldots a_k,a_{k+1})$, in which case $(a_2,\ldots a_k)$ would correspond to 
an internal node. If not, we keep truncating the leading letters of $\bf a$ one by one, until we obtain a 
word $(a_i,\ldots a_k)$ which corresponds to an internal node. Leaf $w$ is then identified with this 
node. The resulting graph for our example is shown in Fig.~2b. 

\begin{figure}
\begin{center}
\psfig{file=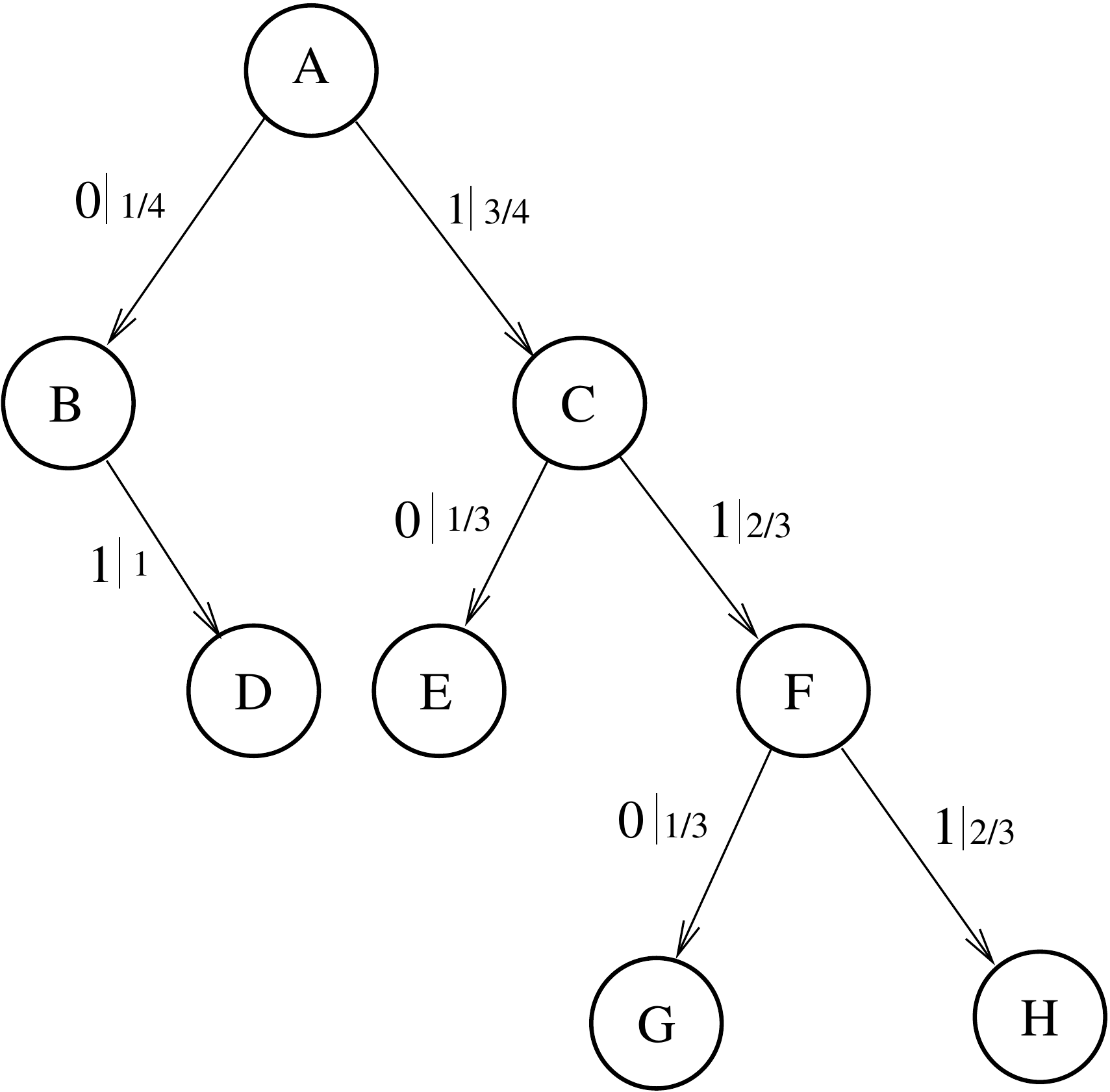,width=5.cm, angle=0}
\vglue .7cm
\psfig{file=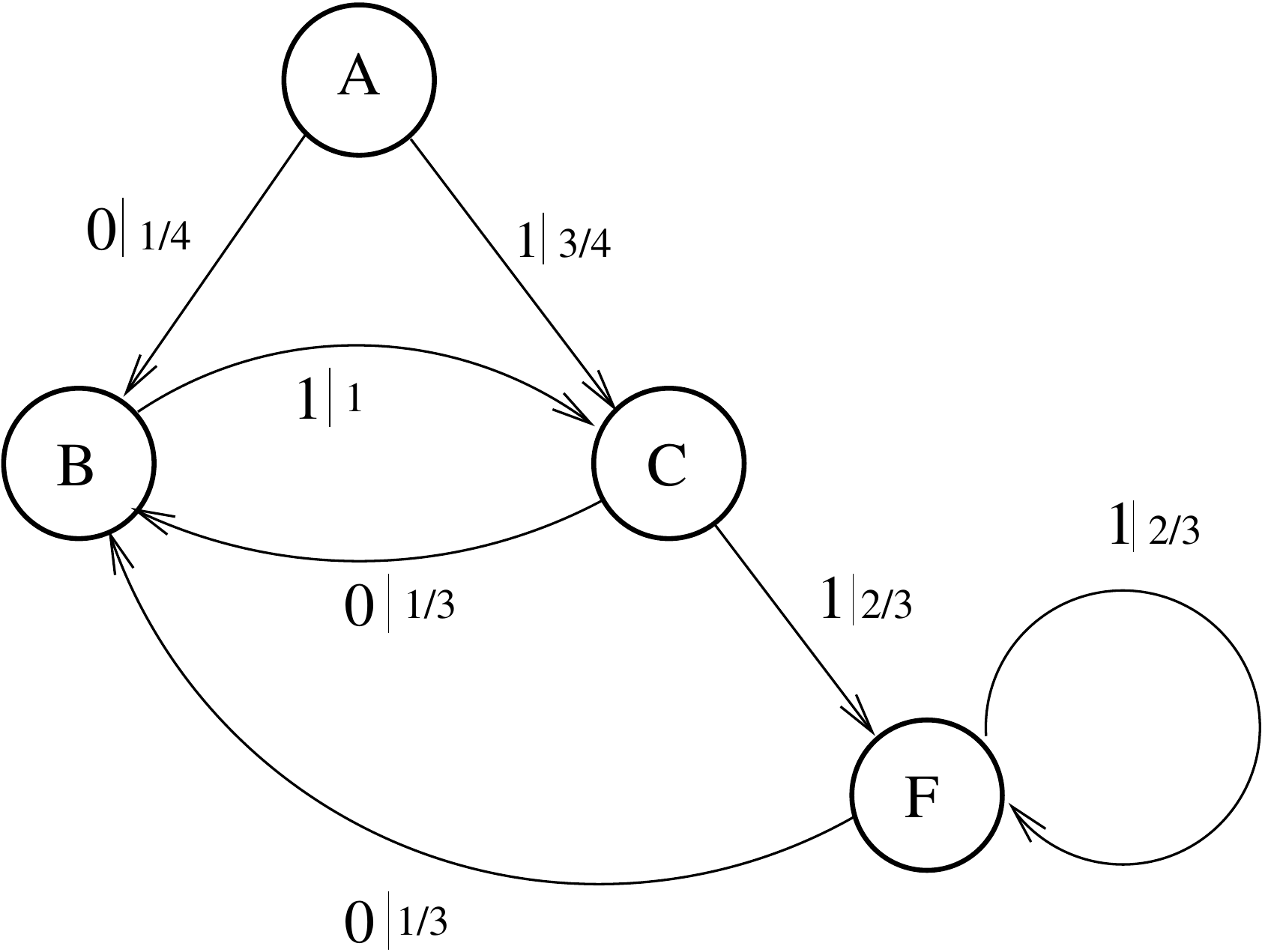,width=5.cm, angle=0}
\vglue .7cm
\psfig{file=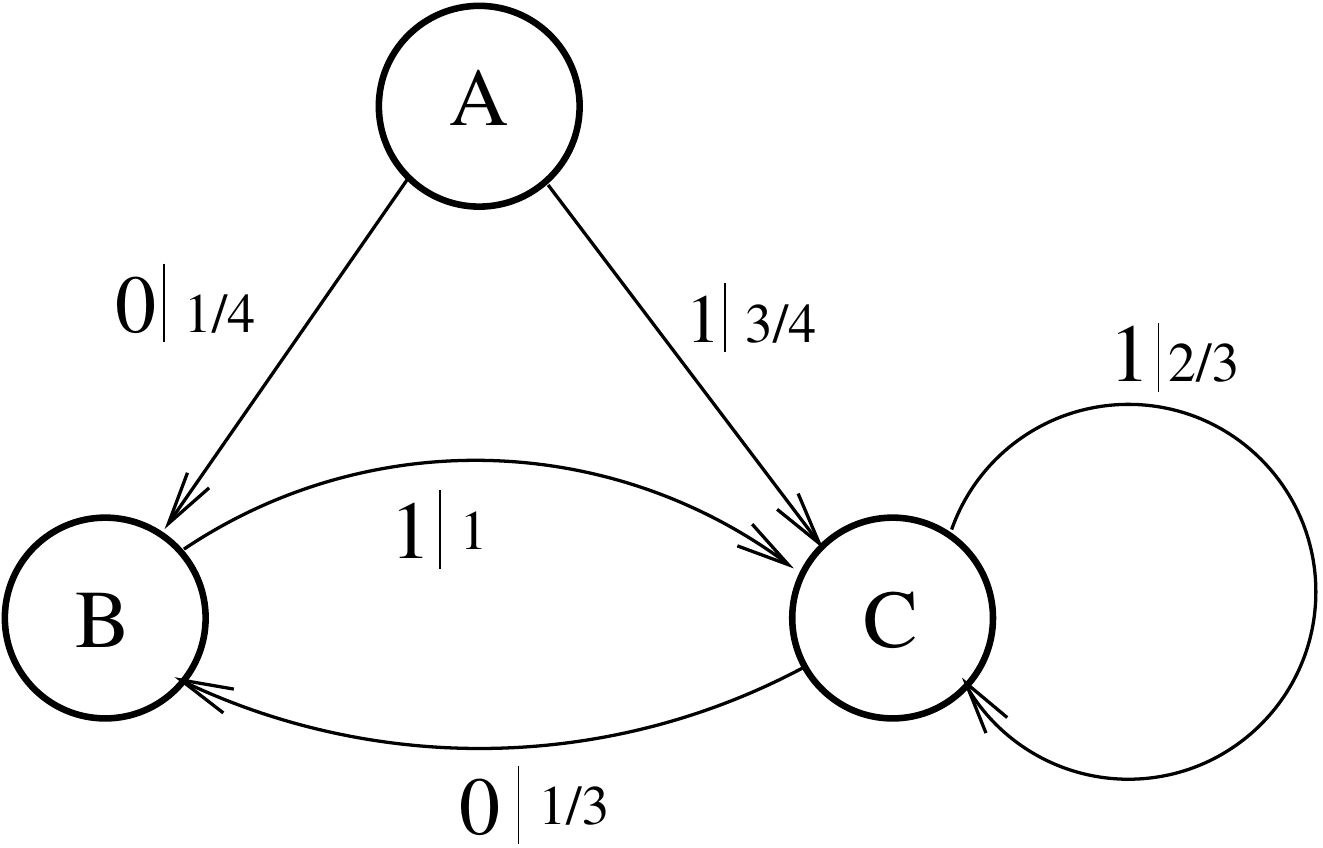,width=5.cm, angle=0}
\caption{(a) Tree obtained from the probabilities given in Eqs.~(1,2). The root will later become 
   the start node.
   (b) Graph obtained by identifying leafs with internal nodes having the same maximal suffixes. 
   (c) Minimal graph obtained by identifying nodes with the same daughters (here: nodes C and F).
   Node A is transient, the other two nodes are recurrent.}
\end{center}
\end{figure}

The last (and most time consuming) step is the simplification by recursively identifying equivalent nodes.
This can be done either divisibly, by first joining nodes into candidate equivalence clusters and then 
splitting these, if they actually contain inequivalent nodes; or agglomeratively, by first assuming all
nodes are inequivalent and then identifying them recursively (the nodes in the minimal graph 
(``$\epsilon$-machine") are called ``causal states" in \cite{Crutch-young1}). An algorithm of the first 
class is described in \cite{Zambella}. Here we describe the agglomerative version.

Two nodes are called equivalent, if they have the same daughters and the same transition probabilities.
In our example, the only equivalent nodes are C and F: They both have the same daughters B and F, and for 
both the branching ratio is 1/3 to 2/3. In general, we have to consider all pairs and identify equivalent 
ones. Once that is done, it might happen that two other nodes have {\it become} equivalent, since two of their
daughters were identified. Thus we repeatedly have to consider again all pairs, until no more equivalence
is found and the number of nodes no longer decreases. Following closely the proof in automata theory 
\cite{hopcroft}, one shows that the result is independent of the order in which pairs are identified 
(the algorithm seems to work fastest,
when one starts with nodes far from the root), and that it is the smallest graph possible. For the present 
example, the result is Fig.~2c. It was easy in all cases to check by hand that no graph with smaller $C$ is 
possible, but I am not aware of any general proof that this gives also the graph with the correct
FC, but it was easy in all cases to check by hand that no graph with smaller FC is possible. In principle 
this is an interesting open problem, but it seems not to be important practically.

\section{Statistical inference of minimal UHMCs; What says Occam's razor really?}

In the above example, the final graph is that for the golden mean process with branching ratio 1:2
at the central node B, including its transient part (node A). This can be truncated if desired,
but we shall argue below that this is usually not a good strategy, and transient parts contain a lot of useful
information. The reason why we could reconstruct the exact graph, although we had only a finite number of 
probabilities to start with, is that the golden mean process is an {\it open} (non-hidden) Markov process.
Thus all the relevant statistics could have been obtained already from words of length 2.

In general, processes are not Markovian. They might be strictly sofic (infinitely many forbidden words, but yet
regular), but they also might be even more complex (with grammars in one of the lower Chomski classes), in which 
case one would even need an infinite graph \cite{Grass86}. In this case, inference from finitely many 
data is of course much more problematic.

There are basically five main problems for inferring a mUHMC from noisy and incomplete data:
\begin{enumerate}
\item A word with low probability might not yet have shown up in the data and thus {\it seems} forbidden, 
although it is not (false negative).
\item A false positive word, i.e. a word which appeared by mistake although it should be forbidden.
\item Two conditional probabilities $\pr\{a_{n+1}|a_n,a_{n-1},\ldots\}$ and $\pr\{a_{n+1}|a_n',a_{n-1}',\ldots\}$
might be very similar and {\it look} the same, although they are not (false equality).
\item The opposite (false inequality).
\item Long forbidden words are not seen. This is different from point \#2 (false positive), in that the 
   occurrence of such words is then inferred by Occam's razor, not on the basis of wrong data.
\end{enumerate}

It was in view of these problems that inference of mUHMCs was deliberately avoided in 
\cite{Grass86,Grass86a,Zambella,Grass88,Grass87,Grass89,Grass89b}.
In retrospect this was an error. The reason is mainly Popper's observation that 
empirical statements can never be verified anyhow: Although we can never, in view of the above list, be 
sure that what we are doing is correct, attempts at statistical inference can nevertheless provide useful
conjectures which should then be subjected to further scrutiny.

Indeed, compared to most other statistical inference problems, the present problem is worse in that
any typical characteristic of the inferred mUHMC (such as the FC ) depends non-continuously
on the input data. More precisely, we expect lower semi-continuity (as proven in \cite{Loehr} for the FC), i.e.
any infinitesimal change of the input data can only induce a jump towards higher complexity, not lower.
Take e.g. the example of Fig.~2, but assume that the probabilities were derived from finite
observations, with the result that $\pr\{1|11\} = 2/3 - \epsilon\pm \delta$ with, say, 
$\epsilon = 3 \delta$. If we (wrongly) assume this difference from $\pr\{1|11\} = 2/3$ to be significant and
thus real, we would not be able to identify the nodes C and F, and the FC would be overestimated by a 
finite amount, irrespective of the magnitude of $\epsilon$. Notice that this could not be improved by 
better data: Then $\epsilon$ would presumably decrease, but so would $\delta$ and we would still be left 
with a finite chance to make the wrong inference.

If we have perfect statistics for all words of length $n$ and no information for longer words, then 
the inferred model will be a Markov process of order $n-1$, as this is the model with largest entropy
compatible with the data \cite{Grass86}. But what if the data do not come from a strictly sofic system
\cite{weiss} or have a non-regular grammar -- or have a finite number of irreducible forbidden words, but 
not a finitely describable probability distribution? In that case the goal should be to reconstruct 
a model with infinitely many forbidden words (resp. probability assignments). A simple example is the 
{\it even process} \cite{Ellison-mahoney-crutch,weiss}, where all subwords of type $0(11)^*0$ are forbidden, 
and all branchings are 1:1. Assume that empirical data are sufficient to construct a Markov approximation 
of order 8. This Markov approximation is given in Fig.~3a. A scientist 
endowed with human intelligence might realize that the correct goal is Fig.~3b, but how can one build 
an {\it algorithm} to do so? In \cite{Shal-Shal-crutch,Shal-shal} the authors do not discuss this 
problem at all, but assume tacitly that a good inference algorithm should make this step. I disagree.
If an inference algorithm finds Fig.~3b instead of Fig.~3a from finite data {\it without being given 
sufficient reason}, then this is a failure and not a success. One cannot expect that a well built 
algorithm should ``guess" what was the intention of the human who set up the problem.

\begin{figure}
\begin{center}
\psfig{file=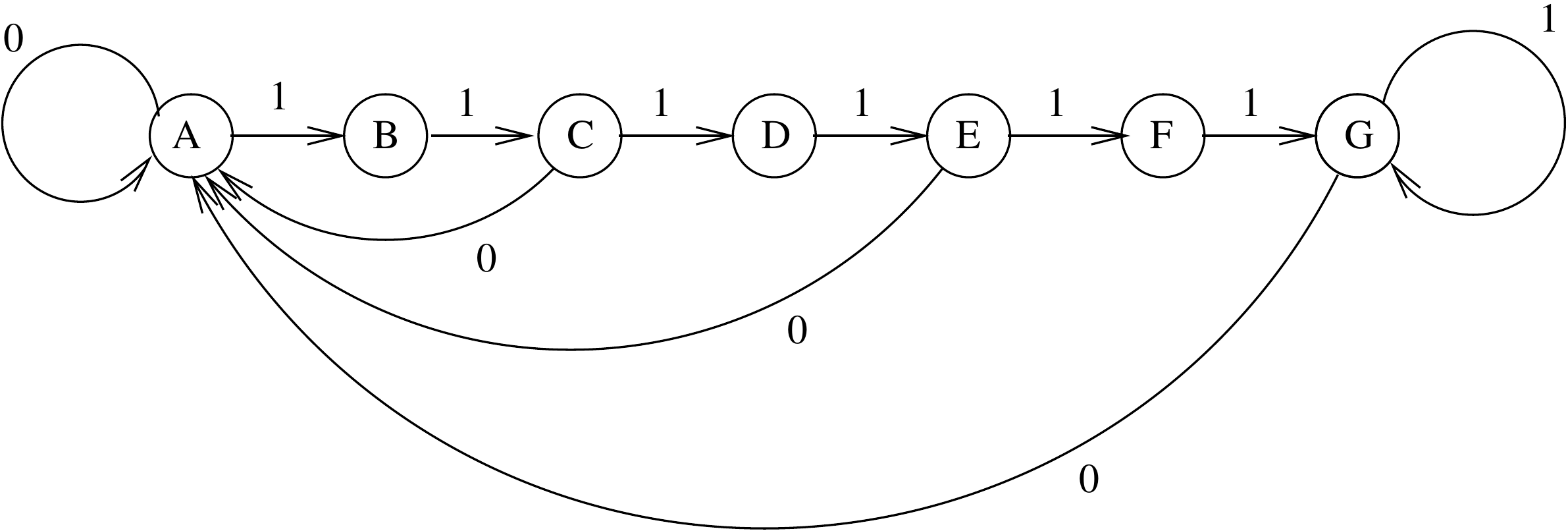,width=7.7cm, angle=0}
\vglue .7cm
\psfig{file=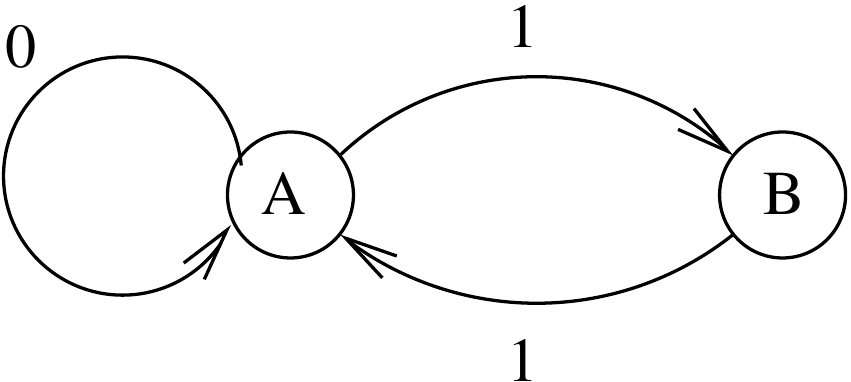,width=4.2cm, angle=0}
\caption{(a) Correct graph obtained from the word probabilities for the `even process' with word lengths $\leq 8$.
   (b) Correct graph for the even process, taking into account {\it all} forbidden words and all 
   word probabilities for arbitrary word lengths. In both graphs, only recurrent parts are drawn.} 
\end{center}
\end{figure}

The point is that the usual application of Occam's razor assumes maximal entropy: There should be 
maximal uncertainty, unless specified otherwise. This is not only the basis of statistical mechanics
\cite{Jaynes}, but also of most of statistical inference. And Fig.~3a definitely has larger entropy
than the ``correct" Fig.~3b.  But yet, the argument that one should 
prefer a {\it structurally} simpler model like Fig.~3b over a structurally less simple one with 
larger entropy is not to be dismissed too easily. Is there any way how we can balance the difference 
in structural information against the difference in entropy?

A reason why this might be possible is that both the entropy $h$ and the forecasting complexity FC
measure difficulties {\it per step of iteration}. Entropy measures the amount of information, and FC
measures the difficulty in using this information \cite{Grass86,Grass89}. If one finds any means of 
weighing these two against each other quantitatively, one might get a handle at a discipled treatment 
of this deep problem. Indeed, this idea is at the basis of Rissanen's {\it minimal description
length} (MDL) \cite{Rissanen}, and of the ``effective complexity" (EC) of Gell-Mann and Loyd                             
\cite{Gell-Mann-lloyd,Lloyd-gell-mann,Ay1,Ay2}. In both cases, however, it was not the sum of FC
and entropy that is minimized, but the total (Shannon / Kolmogorov) information. While the 
informations needed for the ``rules" and for the ``data" are not disentangled in MDL, the EC 
is defined such that the ``rules" are as simple as possible, within well defined error margins
\cite{Ay1}. 

A drawback of the latter approaches is that the two contributions (either entropy and FC, or entropy 
and ``rule" complexity) scale differently with the amount of data available. Thus for different
finite data set, Occam's razor would select a different balance between the two. The practical
difficulties arising from this are best illustrated by the ongoing dispute concerning the AIC and 
BIC model building criteria \cite{Burnham}. 

\section{Causal states do not always correspond to elements of partition; totally recurrent graphs}
    \label{partition}

One central assumption in computational mechanics is that ``causal states" corresponds to elements 
of a partition of state space \cite{footnote8}. We have already seen in the introduction that 
this is not strictly true, since the start node corresponds to the entire state space, thus including 
it would lead to a covering instead of a partition. More generally, nodes in the transient part
of a graph never correspond to partition elements, since the set of all nodes in the recurrent part
cover all state (or history) space (they do, if they do not include the start node, correspond to 
partition elements). But things are actually worse. 

Look at the bottom graph in Fig.~4, taken from Ref.~\cite{Crutch1994} (the top panel of 
Fig.~4 will be discussed later). It accepts all 0/1 sequences in which 
substrings ...00... are forbidden, and where a string $11\ldots 1$ of length $k$ is followed by ``0" 
with probability $k/(2k+2)$. The easiest way to motivate this particular set of probabilities is to 
start with a hidden Bernoulli sequence of letters $A$ and $B$ with $p_A=p_B=1/2$. Subsequent pairs 
of letters are then encoded as $BA \to 0$ and $AA, AB$, and $BB \to 1$. The resulting 0/1 sequence 
is observed, and the minimal graph for optimally forecasting it is the bottom graph in Fig.~4 
(the easiest and most systematic way to obtain such graphs is the method used in \cite{Zambella}
for slightly more complex models where triples of letters are encoded instead of pairs). Node A 
corresponds to the hidden state $A$, while nodes B$_k$ correspond to $A$ and $B$ being in ratio $1:k$.
Node B$_1$, in particular -- corresponding to the stationary ratio $p_A:p_B = 1:1$ -- is the start 
node.

Since the start node does not correspond to a single history but to the entire state space, the same 
is true for all its descendents (since their histories are just extensions of the start node
histories by a finite string of symbols). Thus none of the nodes in this graph corresponds to 
an element of a partition of state or history space.

\begin{figure}
\begin{center}
\psfig{file=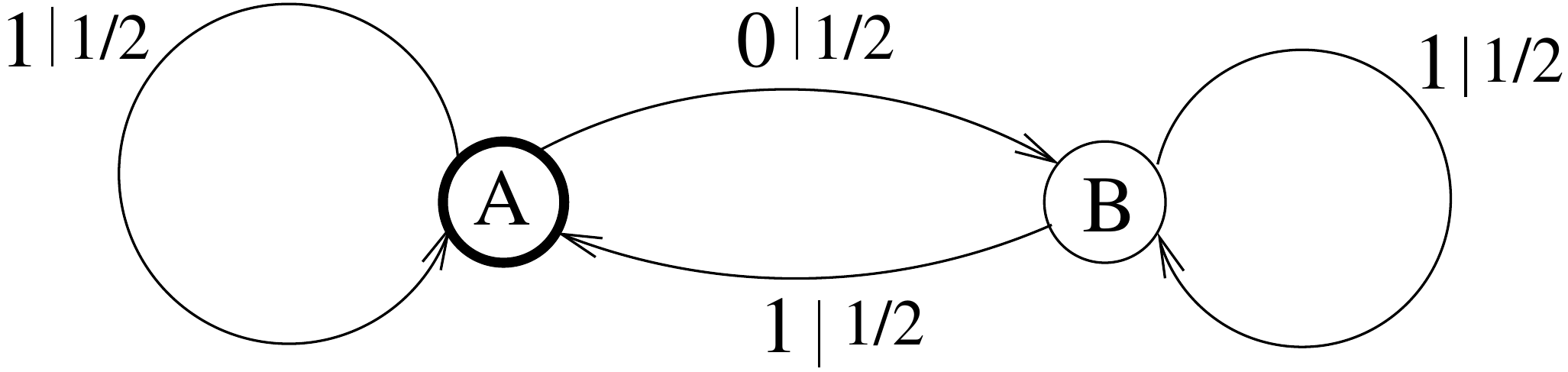,width=5.4cm, angle=0}
\vglue .7cm
\psfig{file=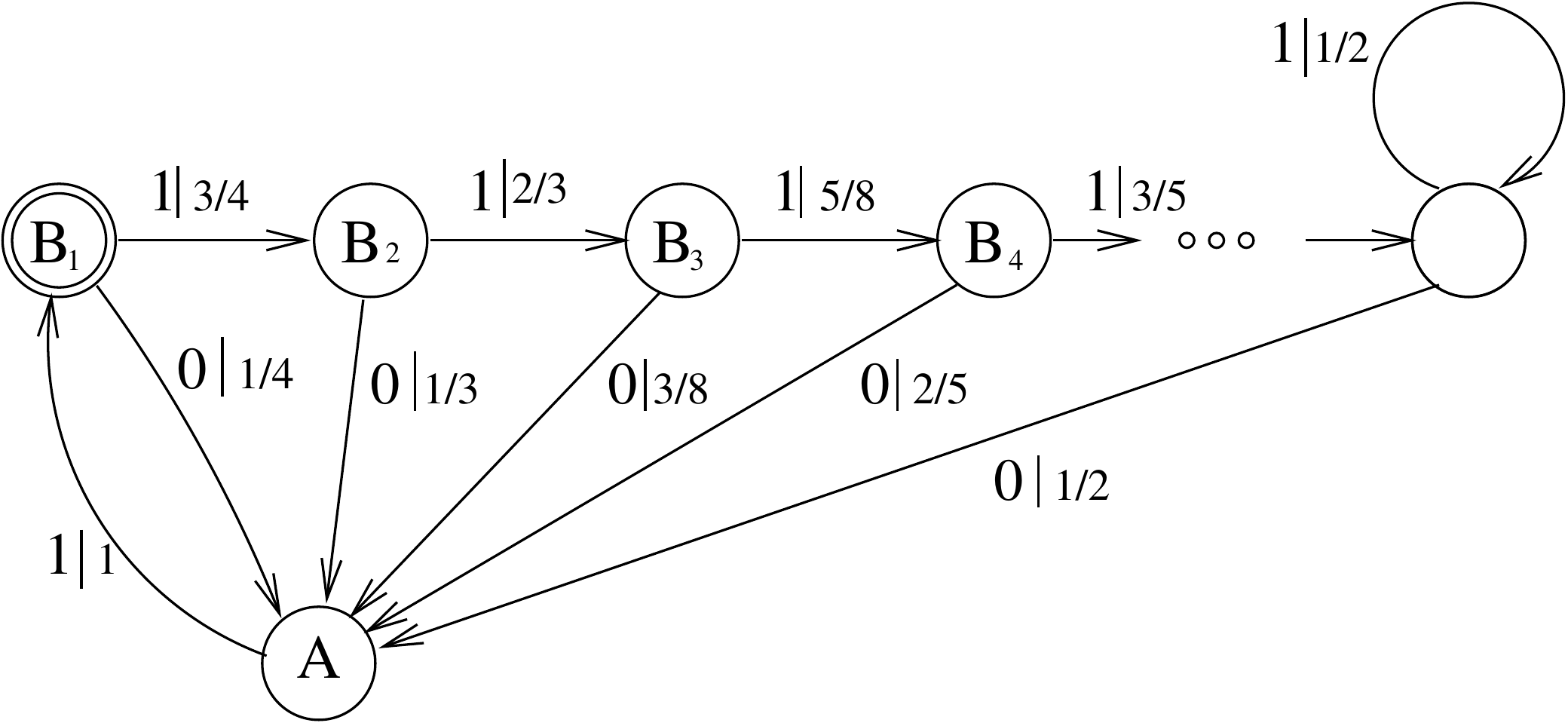,width=7.9cm, angle=0}
\caption{(a) Graph producing sequences without the substring ...00... and with non-trivial
   probabilities for the allowed strings. Notice that two links with label `1' leave node B, i.e. 
   this graph is {\it not unifilar}. Node names are arbitrary.
   (b) Part of the minimal infinite graph that produces strings which have both the same grammar 
   and the same probabilities as the graph in panel (a). 
   In graph (b), the heavy node is the start node. In graph (a), the heavy node can be used as
   start nodes for verifying grammatical correctness. For optimal forecasts from the very beginning on,
   one would need a transient part which is omitted for simplicity
   (modified from \cite{Crutch1994}).}
\end{center}
\end{figure}

A similar case had been given before in \cite{Zambella}, where mUHMC's were studied that were 
obtained by encoding triples in $A/B$ Bernoulli strings in terms of 0 and one. The model called 
R22 in \cite{Zambella} was e.g. defined by mapping $AAA,ABB,BAB,BBA,BBB \to 0$ and $AAB,ABA,BAA \to 1$
(the model in Fig.~4. was called R12 in \cite{Zambella}).
This model also gave rise to an infinite mUHMC graph, but of much more complex structure. It 
does not have a single branch extending to infinity but infinitely many. Nevertheless we could prove
that it had a finite forecasting complexity due to the ``resetting string" 00001101000010000. This
word appears with probability one in the future history of any node, and when it has appeared the 
walker on the graph is re-set to the start node. This is completely analogous to Fig.~4b, where 
the resetting string is 01.  

We propose to call mUHMC's where the start node (and thus also all other nodes) is recurrent
``totally recurrent" graphs. Although we do not have a proof for it, it seems that all totally
recurrent graphs are infinite \cite{footnote6}. Finite graphs with supposedly recurrent start nodes 
were shown in Figs.~14b to 14d in \cite{Crutch1994}, but they are all wrong.

The fact that nodes in mUHMC's do not always define a partition of history space but do in general
define a {\it covering} had first been pointed out in \cite{Grass88}.

\section{Why is unifilarity needed?}

In \cite{Loehr-ay} the authors observe that model complexity (as measured by the FC) can in general be
reduced, if the assumption of unifilarity is dropped. They present an explicit example (Example 3.6) where
the number of nodes is finite for a non-unifilar machine (Fig.~6 in \cite{Loehr-ay}), but any unifilar
HMC would need an infinite number of nodes. Based on this, they develop a theory of optimal prediction
which is more general than computational mechanics and uses a forecasting device which is supposedly
more powerful than ``$\epsilon-$machines" (i.e. mUHMCs).

Indeed, a similar but somewhat simpler example is provided by Fig.~4. That example is sufficient
to illustrate the main aspects of this problem. Its most simple non-unifilar representation is given
in Fig.~4a. It was justly rejected in \cite{Crutch1994}, but for wrong reasons. We shall therefore
discuss it again in some detail, in order to show why unifilarity is important.

The Shannon information $C$ needed at any given time to locate a walker on Fig.~4a, and the information
$h_a$ needed to identify the next step, are equal to 
\be
    C = 1\; {\rm bit},\;\;\; h_a = 1\; {\rm bits}.
\ee
At the same time, forecasting such walks gives also optimal forecasts for the 0/1 symbol string. Thus 
probabilities of symbols can be forecasted perfectly and optimally from Fig.~4a (after it has been
augmented eventually by its transient part; but this does not affect FC), in contrast to what is said
in \cite{Crutch1994}.

On the other hand, 
forecasting complexity and entropy of Fig.~4b --  which are indeed the correct values for the 0/1 
symbol string -- are easily calculated to be \cite{Zambella,Crutch1994}
\be
    FC = 2.7114687...\;\; {\rm bits},\;\;\; h = 0.67786718...\; {\rm bits}.
\ee 

This should make clear what is wrong with Fig.~4a. While it has a smaller structural complexity than
the unifilar machine in Fig.~4b, it needs more information to make its forecasts. The difference in 
entropy corresponds to the fact that
the machine shown in Fig.~4a does {\it more} than just predict the 0/1 symbol sequence: It predicts
{\it also} the present vertex of the graph. Since the states of non-unifilar hidden Markov models are 
not unique functions of the symbol sequence (that is why they are also called ``non-deterministic" in 
automata theory: the determinism meant here is not that of the symbol sequence, but that of 
the machine, given the symbol sequence), one needs more information for specifying the sequence of 
machine states than for specifying the symbol sequence alone.

This makes it now clear what singles out mUHMCs among all hidden Markov chains: They allow prediction 
using the smallest possible amount of information per letter. It is this requirement -- which essentially 
demands that the input information is used {\it only} for forecasting the sequence, and is not wasted on 
other tasks -- that makes the entire enterprise meaningful. One might try to strike some balance between
structural complexity and entropy, as discussed at the end of the last section, but this is subtle and 
would have to be done carefully.

Notice also that this has nothing to do with the information bottleneck treated in \cite{Still}. In the 
latter, one poses the problem how much one wins in reduced complexity by allowing non-optimal forecasts,
which do not use more than the minimal information. This is more closely related to the inference 
problem discussed in the last section.
Using non-unifilar hidden Markov chains would allow one still to make optimal forecasts at reduced 
complexity, but at the cost of increased information input.  

\section{Transient states and computation of the excess entropy}

One basic feature which distinguishes computational mechanics from conventional Markov chain theory
is the explicit treatment of transient parts of the Markov diagram. This is inherited from the 
theory of finite automata from which the basic concepts were borrowed in \cite{Grass86}. Transient 
states were shown regularly in the earlier papers by Crutchfield {\it et al.} 
\cite{Crutch-young1,Crutch-young2,Crutch1992,Crutch1994}, although they contradicted one of the 
main assumptions on which these papers are supposedly based: That states in ``$\epsilon$-machines"
correspond one-to-one to elements of {\it partitions} of the space of past histories. This assumption 
is not true for transient states, but -- as we have seen -- in some ``$\epsilon$-machines" it is not 
true for {\it any} node.

In recent papers on ``computational mechanics", transients no longer are discussed. This might be 
because authors have realized that they do not correspond to partition elements, but it was never
said explicitly to my knowledge. Anyhow this is a pity, because transient parts carry extremely 
useful information. On the one hand, there are processes (such as the set of all trajectories on the
Feigenbaum attractor) where any finite part is transient, so nothing could be done without transient 
parts. But also in other cases the transient part is crucial to predict during initial phases of 
observation. Finally, it is useful even for problems which seem to deal only with stationary 
situations. To illustrate this we discuss the calculation of the excess energy {\bf E}. In 
\cite{Crutch-ellison-mahoney} it is stated that ``To date, {\bf E} cannot be as directly calculated 
or estimated as the entropy rate and the statistical complexity." The authors then go on and 
propose a new ingenious method to compute it, given the recurrent parts of both the forward and
backward Markov diagrams. They ignored that a very simple (much simpler than theirs) 
method using the full Markov diagram had been proposed twenty years earlier in \cite{Zambella}. 

Consider an mUHMC with nodes $S_i$ and probabilities $T_{ki}^{(a)}$ for transiting from node $i$ to 
node $k$ by emitting symbol $a$. In the following we will assume for simplicity that $a\in\{0,1\}$. 
The start node is $S_0$. Any (normalized) probability distribution
on the nodes will be denoted as $p =\{p_i\}$ where $p_i$ is the probability to be at node $i$. 
Consider now a distribution which is concentrated at time $n=0$ on the start node, $p_i^{(0)} =
\delta_{i,0}$. At subsequent times $p$ follows the master equation
\be
    p_i^{(n+1)} = \sum_k \sum_{a_n} T_{ik}^{(a_n)}p_k^{(n)}.             \label{p}
\ee
Iterating this master equation allows one also to compute the block entropies
\be
   H_n = -\sum_{a_0\ldots a_{n-1}} \pr\{a_0\ldots a_{n-1}\} \log \pr\{a_0\ldots a_{n-1}\}
\ee
and their increments
\ba
   h_n & \equiv & H_{n+1}-H_n   \\ \nonumber
    & = & -\sum_{a_0\ldots a_n} \pr\{a_0\ldots a_n\} \log \pr\{a_n|a_0\ldots a_{n-1}\}. \label{hn0}
\ea
Indeed, $\pr\{a_n|a_0\ldots a_{n-1}\}$ is just the transition probability $T_{ki}^{(a_n)}$ evaluated
at the node reached by the word $a_0\ldots a_{n-1}$, and summing over all trajectories of $n$ steps
ending at node $i$ gives
\be
   \sum_{a_0\ldots a_{n-1}: S_0\to i} \pr\{a_0\ldots a_{n-1}\} = p_i^{(n)}.
\ee
Therefore \cite{Zambella}, 
\be
   h_n = -\sum_i\sum_{a} p_i^{(n)} T_{ki}^{(a)} \log T_{ki}^{(a)}      \label{hn}
\ee
and
\be
   h = \lim_{n\to\infty} h_n = -\sum_i\sum_{a} p_i^{(\infty)} T_{ki}^{(a)} \log T_{ki}^{(a)}\;,      \label{h}
\ee
which gives finally
\ba
   {\bf E} & = & \sum_{n=0}^\infty (h_n - h)       \\ \nonumber
           & = & \sum_{n=0}^\infty \sum_{i_n}\sum_{a_n} (p_{i_n}^{(\infty)}-p_{i_n}^{(n)}) T_{ki}^{(a_n)} \log T_{ki}^{(a_n)}\;.
\ea
Notice that the most precise calculation of $h$ also proceeds
in general via Eq.~(\ref{h}) \cite{Gallager,Zambella,dalessandro}.

Analogously to {\bf E}, a ``transient information" was defined in \cite{Crutch-feld03} as
\be
   {\bf T} = \sum_{n=0}^\infty (n+1)(h_n-h).
\ee
The above way of computing $h_n$ helps of course also in computing ${\bf T}$ much more 
precisely. As an application we mention the ``RRXOR" process (\cite{Crutch-feld03}, sec.~VI.D).
While values ${\bf E} = 2$ bits and ${\bf T} \approx 9.43$ bits are quoted in \cite{Crutch-feld03},
the exact values are ${\bf E} = 2.2516...$ bits and ${\bf T} = 12.743...$ bits. One reason why
${\bf E}$ and ${\bf T}$ are so big for the RRXOR process is that it has 31 transient nodes (and 
only 5 recurrent ones), and the $h_n$ converge very slowly, $h_n - h \sim 2^{-n/3}$ 
(the estimate in \cite{Crutch-feld03} was $h_n-h\sim 2^{-0.306(1)n}$).

\section{``$\epsilon-$machines"}

The diligent reader will have realized that I have avoided the name ``$\epsilon-$machines" as 
far as possible. First of all, they were also introduced in \cite{Grass86}, where they were called 
``minimal deterministic automata". But more importantly, 
the ``$\epsilon$" in this name originated historically from the assumption in \cite{Crutch-young1}
that one has to start by making first a partitioning of phase space into a grid with mesh size
$\epsilon$. The idea was then that this introduces some small error (if $\epsilon$ is small), 
whose effect on subsequent results has to be studied carefully. Together with another small 
coarse graining parameter, $\delta$, which controls tolerable differences in transition 
probabilities, this would allow a detailed scaling theory of statistical inference for real
valued phase spaces.

This would have been a formidable program, but it was never realized. Some (not very systematic) 
studies in this direction had been made already before in \cite{Zambella}, but that is to my 
knowledge all. Anyhow, a better strategy than making explicit $\epsilon-$partitions is to use  
symbolic dynamics based on generating partitions, as e.g. in 1-d one-humped maps. Unfortunately, 
in higher dimensions exact generating partitions are known in very few cases only. In view of this,
in all papers on computational mechanics I am aware of the authors studied 1-d maps, or a discrete
phase space was assumed from the very beginning. Thus the ``$\epsilon$" in ``$\epsilon-$machines"
stands now for nothing.

In view of this I propose to call them ``forecasting graphs" instead, because this name is simpler 
than the one used originally in \cite{Grass86}, and they are precisely
what is needed for optimal forecasts.

\section{Conclusions}

These notes grew out of increasing bewilderment and anger about the literature on computational
mechanics. I found that concepts and results which I and my collaborators had clearly developed 
more than twenty years ago were explicitly attributed to later authors. As I read more and more 
of this literature, my bewilderment slightly shifted. Not all of our results were copied, some 
rested undiscovered, although sub-optimal or even wrong attempts (like e.g. the reconstruction of 
Markov graphs from frequency data) flourished. And some misconcepts were perpetuated, although
the correct solutions were in principle available in the literature.

But after starting to write these nodes I realized that there is something which might be much 
more interesting for other readers than my personal troubles. By putting things straight, 
several new and interesting avenues opened which neither I nor others seem to have anticipated.
I hope that these positive achievements might make the present notes of interest even to a 
wider readership.

I am indebted to correspondence to Philippe Binder, Cosma Shalizi, Karoline Wiesner, and 
Wolfgang L\"ohr. Even more so I am indebted to Chris Ellison and John Mahoney for discussions.

\section{Note added (April 3, 2018)}

In arXiv:1710.06832, Dr. Crutchfield published a rebuttal of the present comment. 

In this rebuttal, he makes very general statements like 

\begin{itemize}

\item ``[...] misguided evaluations of [...] 
computational mechanics are groundless and stem from a lack of familiarity with its basic
goals and from a failure to consider its historical context";

\item ``[...] its modern methods and results largely supersede the early works";

\item ``[...] renders recent criticism moot [...]";

\item ``[...], they are in large measure misguided";

\item ``They are based on arguments that selectively pick details, either by quoting them out 
of context or applying inappropriate contexts of interpretation",

\end{itemize}

but he actually refutes {\it none} of the numerous very precise points I raised.

There are actually occasions where he became more specific, but also there 
he did not really show that I am wrong, although his words suggest so:

\begin{itemize} 

\item On page 5, he enters a long discussion about ``Effective states and equivalence relations", 
with the implicit suggestion that I had not been aware of these. But of course I had been aware,
and he actually had first heard from me about them. 
In \cite{Grass86} I had developed the theory largely as an extension of the theory of formal 
languages (where sequences are characterized by grammatical rules, but with no probability measures
associated to them), and I could thus be rather short on concepts like minimal deterministic automata, 
causal states, automaton reconstruction algorithms, equivalence between automata, etc., because they 
are largely the same in the theories with and without probability measures.

\item Also on page 5, he mentions `elusive' and `unreachable' states that I had supposedly missed in
\cite{Grass86}, and that would render my arguments obsolete. But I had not missed them. Instead,
I had stated explicitly that I consider only sequences where all symbols occur with non-zero
frequency (which excludes such states). 

\item In the same paragraph, he cites papers on what he calls synchronization. I had not mentioned
them, from which he obviously concluded that I did not know them (otherwise I should have 
criticized them?)

\item At the end of Sec.2 I said that his claim for having invented `excess entropy' {\bf E} is wrong,
because it was introduced by Shaw \cite{Shaw} (of which I was not aware when I called it `effective measure
complexity' in \cite{Grass86}).
In \cite{Crutch-packard}, Packard and Crutchfield had indeed called `excess entropies' the differences
$h_n - h$ between the increments $h_n$ (defined in Eq.(\ref{hn0})) and the entropy $h=\lim_{n\to\infty}$.
But they did not mention {\bf E} (which is called excess entropy now), as they obviously were not aware
of its central r\^ole as being the mutual information between past \& future and a lower bound to the 
forecasting complexity \cite{Grass86}).
On page 7 of his rebuttal, Crutchfield now 
says that {\bf E} was already used previously in Norman Packard's thesis (which I have to believe), 
that Shaw's results were actually obtained by Shaw and himself (which I might also believe), and that  
I had completely misunderstood \cite{Crutch-packard}. The last is definitely not true -- anybody is 
invited to read that paper.

\end{itemize}

In his rebuttal, Dr. Crutchfield discusses at length his excellent academic background, his great early 
work on deterministic chaos, and his important recent contributions to what he calls `computational
mechanics'. I do not deny any of these, but I don't know why they are relevant to the present issue.
This issue is that 

\begin{itemize} 

\item
Although he had cited my paper \cite{Grass86} in his very first paper with K. Young \cite{Crutch-young1}
(which proves that he was aware of my work), he later claimed that he had invented most of 
my results -- including forecasting complexity (which he renamed `statistical complexity'), 
the minimal forecasting graphs (which he renamed `$\epsilon$-machines'),
and their nodes (which he renamed `causal states'). 

\item He never cited my subsequent papers, although they would have prevented him from many mistakes and
they contain results that are still relevant today:
\begin{itemize}

\item In \cite{Zambella} we gave an algorithm for computing {\bf E}
which is still the most efficient and simple algorithm for this purpose.
\item There we also gave the best (up to now) algorithm for estimating optimal forecasting graphs 
from incomplete data.
\item There and in \cite{Grass88} we proved that `causal states' do not always correspond to
elements of a state (or history) state {\it partition}, but can correspond in some cases to elements
of a covering where states are covered more than once. The opposite is still claimed
in all the recent pertinent literature.
\end{itemize}

\item Why should one use a non-descriptive term like `statistical complexity', if the precise notion
`forecasting complexity' had been used by the inventor of the concept, and was used by him already in
several papers?

\end{itemize}

As a side remark, a supposed `$\epsilon$-machine' for the Feigenbaum attractor
which had been for nearly 20 year on his homepage, disappeared after I had pointed it out
to be wrong.  Maybe, not all of my points are so moot. 

On page 7 he says ``The nontechnical and ad hominem criticisms intertwined with the technical faults 
are evidence of the consistent projection of irrelevant meanings onto the material", and claims 
that my claims are ``based on false memories". But first of all, 
he could not point at a single technical fault. Secondly, everyone can read e.g. \cite{Grass86}
and \cite{Crutch-young1}, verifying thereby the `false memories'. And finally, it was not me who 
lied bluntly about where the foundations of `computational mechanics' are to be found, even 
after I had tried to correct history in more friendly ways. I guess that any reply to such behavior 
then is ad hominem.


\begin{thebibliography}{99}
\bibitem{footnote1} A hidden Markov model has internal (hidden) states which can be represented 
   as nodes of a directed graph. Associated to each link (transition) from node $A$ to $B$ is 
   the emission of an (observed) output symbol $a$. A hidden Markov chain is called unifilar, if 
   the pair $\{A,a\}$ determines $B$ uniquely, i.e. if all links from some hidden node $A$ to any 
   other node are associated with different symbols. In automata theory, this is usually called 
   ``determinism" of the automaton, not to be confused with determinism of the underlying process.
   Unifilar hidden Markov chains were called ``Markov sources" in \cite{Gallager}.
\bibitem{footnote9} One technical difficulty is that minimal graphs are obtained iteratively, by 
   either joining or splitting nodes. The order of these actions is not fixed. One can show \cite{hopcroft}
   that the size of the minimal graph is unique, but not necessarily the graph itself. Different minimal
   graphs might have different forecasting complexities. This difficulty is aggravated in cases where
   the minimal graph is infinite.
\bibitem{footnote8} More precisely, in \cite{Shalizi-crutch} it is claimed that they correspond to 
   elements of a partition of the space of all histories. But in the spirit of delay embeddings,
   histories can always be considered as generalized coordinates, so that a partitioning of 
   history space can also be considered a a partitioning of state space.
\bibitem{footnote6} The same is not true for automata verifying {\it grammatical} correctness only. There,
   if one recurs to the start node, one can do this with different probability from the previous 
   recurrence. In mUHMC's, this would correspond to two different nodes.
\bibitem{footnote3} An excellent test for the correctness of the reconstructed part consists in 
   calculating the entropies $h_n$ by Eq.~(7). They should form a monotonically decreasing sequence. 
   For sloppy attempts to reconstruct transients manually, this was usually not the case.
\end{thebibliography}
\end{document}